\documentclass[twocolumn]{aastex62}
\usepackage{array,url,kantlipsum}
\usepackage{booktabs}
\usepackage{verbatim} 
\usepackage{natbib}

\newcommand{\AS}{AS\,205\,N }
\newcommand{\HT}{HT\,Lup\,A }

\graphicspath{{./}{figures/}}

\shorttitle{DSHARP Multiple Systems}
\shortauthors{Kurtovic et al.}

\begin{document}

\title{The Disk Substructures at High Angular Resolution Project (DSHARP): \\
IV. Characterizing substructures and interactions in disks around multiple star systems.}

\correspondingauthor{Nicol\'as T. Kurtovic}
\email{nicokurtovic@gmail.com}

\author{Nicol\'as T. Kurtovic}
\affil{Departamento de Astronom\'ia, Universidad de Chile, Camino El Observatorio 1515, Las Condes, Santiago, Chile}

\author[0000-0002-1199-9564]{Laura M. P\'erez}
\affil{Departamento de Astronom\'ia, Universidad de Chile, Camino El Observatorio 1515, Las Condes, Santiago, Chile}

\author[0000-0002-7695-7605]{Myriam Benisty}
\affil{Departamento de Astronom\'ia, Universidad de Chile, Camino El Observatorio 1515, Las Condes, Santiago, Chile}
\affiliation{Unidad Mixta Internacional Franco-Chilena de Astronom\'{i}a, CNRS/INSU UMI 3386}
\affiliation{Univ. Grenoble Alpes, CNRS, IPAG, 38000 Grenoble, France.}

\author[0000-0003-3616-6822]{Zhaohuan Zhu}
\affiliation{Department of Physics and Astronomy, University of Nevada, Las Vegas, 4505 S. Maryland Pkwy, Las Vegas, NV, 89154, USA}

\author[0000-0002-8537-9114]{Shangjia Zhang}
\affiliation{Department of Physics and Astronomy, University of Nevada, Las Vegas, 4505 S. Maryland Pkwy, Las Vegas, NV, 89154, USA}

\author[0000-0001-6947-6072]{Jane Huang}
\affil{Harvard-Smithsonian Center for Astrophysics, 60 Garden Street, Cambridge, MA 02138, USA}

\author{Sean M. Andrews}
\affil{Harvard-Smithsonian Center for Astrophysics, 60 Garden Street, Cambridge, MA 02138, USA}

\author[0000-0002-7078-5910]{Cornelis P. Dullemond}
\affil{Zentrum f\"r Astronomie, Heidelberg University, Albert Ueberle Str. 2, 69120 Heidelberg, Germany}

\author[0000-0002-0786-7307]{Andrea Isella}
\affiliation{Department of Physics and Astronomy, Rice University 6100 Main Street, MS-108, Houston, TX 77005, USA}

\author[0000-0003-1172-3039]{Xue-Ning Bai}
\affil{Institute for Advanced Study and Tsinghua Center for Astrophysics, Tsinghua University, Beijing 100084, China}

\author[0000-0003-2251-0602]{John M. Carpenter}
\affil{Joint ALMA Observatory, Avenida Alonso de C\'ordova 3107, Vitacura, Santiago, Chile}

\author[0000-0003-4784-3040]{Viviana V. Guzm\'an}
\affil{Joint ALMA Observatory, Avenida Alonso de C\'ordova 3107, Vitacura, Santiago, Chile}
\affil{Instituto de Astrof\'isica, Pontificia Universidad Cat\'olica de Chile, Av. Vicu\~na Mackenna 4860, 7820436 Macul, Santiago, Chile}

\author{Luca Ricci}
\affil{Department of Physics and Astronomy, California State University Northridge, 18111 Nordhoff Street, Northridge, CA 91130, USA}

\author[0000-0003-1526-7587]{David J. Wilner}
\affil{Harvard-Smithsonian Center for Astrophysics, 60 Garden Street, Cambridge, MA 02138, USA}

\begin{abstract}

To characterize the substructures induced in protoplanetary disks by the interaction between stars in multiple systems, we study the $1.25\,$mm continuum and the $^{12}$CO$(J=2-1)$ spectral line emission of the triple systems HT\,Lup and AS\,205, at scales of $\approx 5\,$au, as part of the ``Disk Substructures at High Angular Resolution Project'' (DSHARP). 
In the continuum emission, we find two symmetric spiral arms in the disk around AS\,205\,N, with pitch angle of $14^\circ$, while the southern component AS\,205\,S, itself a spectroscopic binary, is surrounded by a compact inner disk and a bright ring at a radius of $34\,$au. 
The $^{12}$CO line exhibits clear signatures of tidal interactions, with spiral arms, extended arc-like emission, and high velocity gas, possible evidence of a recent close encounter between the disks in the AS\,205 system, as these features are predicted by hydrodynamic simulations of fly-by encounters.
In the HT\,Lup system, we detect continuum emission from all three components. 
The primary disk, HT\,Lup\,A, also shows two-armed symmetric spiral structure with a pitch angle of $4^\circ$, while HT\,Lup\,B and C, located at $25$ and $434\,$au in projected separation from HT\,Lup\,A, are barely resolved with $\sim5$ and $\sim10\,$au in diameter, respectively. 
The gas kinematics for the closest pair indicates a different sense of rotation for each disk, which could be explained by either a counter rotation of the two disks in different, close to parallel, planes, or by a projection effect of these disks with a close to $90^\circ$ misalignment between them.

\end{abstract}

\keywords{protoplanetary disks, stars: binaries (close), ISM: dust, techniques: high angular resolution}



\section{Introduction} \label{sec:intro}

In the early stages of star formation, the conservation of angular momentum through the gravitational collapse leads to the formation of a gas and dust disk around the young forming star; it is here where planetary systems may form. Given that most stars live or appear to have been formed in binary or multiple systems \citep{Raghavan2010, Duchene2013}, it is expected that companions or close encounters will modify the disks in multiple stellar systems, when compared to disks around single, isolated stars.

Nonetheless, planets have been detected around single stars in multiple systems \citep[e.g.,][]{eggenberger2007,chauvin2011}, mostly at separations larger than few tens of au, although this might be an effect of selection biases \citep{winn2015}. Circumbinary planets have also been found \citep{doyle2011}, sometimes in systems with more than one planet \citep[Kepler-47,][]{kostov2013}. However, \citet{wang2014} find that planets should be 4.5$\pm$3.2 and 2.6$\pm$1.0 times less frequent in a multiple system (compared to single star systems), when a stellar companion is at a distance of 10 and 100\,au, respectively  \citep[see also][]{Kraus2016}.

Over the last few years, the detection and characterization of dust and gas structures in disks at high angular resolution has been helping us to understand the processes involved in the evolution of young stars and the formation of planetary systems. Millimeter images that trace dust emission have shown large azimuthal asymmetries \citep[e.g.,][]{vanderMarel2013}, spirals \citep[e.g.,][]{Perez2016}, and bright/dark rings \citep[e.g.,][]{ALMAPartner2015}, the latter appear to be the most common substructure in Class II disks \citep{Andrews2018,Huang2018a}. These features are often interpreted as signposts of planet-disk interactions \citep[e.g.,][]{zhu2011}. 

While multiplicity is high among young stars \citep[e.g., $\sim0.7$ in Class II/III stars, $\sim 0.6$ in Class 0 objects,][]{Kraus2011,Tobin2016b}, most observational studies at high angular resolution have so far focused on single stellar systems, and it is not clear how common such disk substructures are in multiple systems. 
Over the last two decades, a few multiple systems have been studied with sufficient angular resolution to resolve their components separately at radio wavelengths \citep{Jensen1996, Jensen2003, Harris2012}, and thanks to ALMA capabilities these detections have become more common over the last few years \citep[e.g.,][]{Akeson2014, Jensen2014, Brinch2016, Tobin2016}. 
Misalignments between the disk rotation axis and the binaries orbit \citep{Williams2014,Fernandez2017}, as well as tidal stripping and extended emission \citep{Cabrit2006, Salyk2014, Rodriguez2018} have been detected in some of them. However, none of the previous observations reached the high spatial resolution required to study the  substructure of disks in multiple systems.
For example, spiral arms are expected to be triggered in the presence of one or multiple companions \citep{Goldreich1979, Tanaka2002}, and could potentially be detected in the dust continuum emission at millimeter wavelengths. However, until today, most objects with known spiral-like structures in dust continuum emission \citep{Perez2016, Dong2018, Boehler2018} or gas emission \citep[e.g.,][]{Tang2017} do not host any known stellar companion \citep[an exception is HD\,142527, ][]{Biller2012,Christiaens2014,Christiaens2018}.
This suggests that in these cases, the observed spirals  might originate from other mechanisms, such as gravitational instabilities \citep{Mayer2004, Lodato2004}, shadowing in the disk \citep{Montesinos2016}, or alternatively, that the possible companion(s) have not yet been detected.

The present work is the first step towards the detection and characterization of disk substructures, with high angular resolution observations ($\sim 5$\,au) of multiple systems, which will help us understand how stellar interactions affect the evolution of gas and dust in protoplanetary disks. We present the first analysis of two young multiple systems, HT\,Lup and AS\,205, observed as part of our ALMA Large Program ``DSHARP: Disk Substructures at High Angular Resolution Project'' \citep{Andrews2018}. In Sect.~\ref{sec:targets}, we present the targets, in Sect.~\ref{sec:data}, we briefly describe the observations and specific calibration and imaging procedures. In Sect.~\ref{sec:results}, we present our analysis and modeling of these new data, that we further discuss in Sect.~\ref{sec:discussion}. Finally we conclude and summarize our results in Sect.~\ref{sec:conclusions}.


\section{Targets} \label{sec:targets}

\subsection{AS 205} \label{sec:as205}

\begin{figure*}[ht!]
\includegraphics[width=\linewidth]{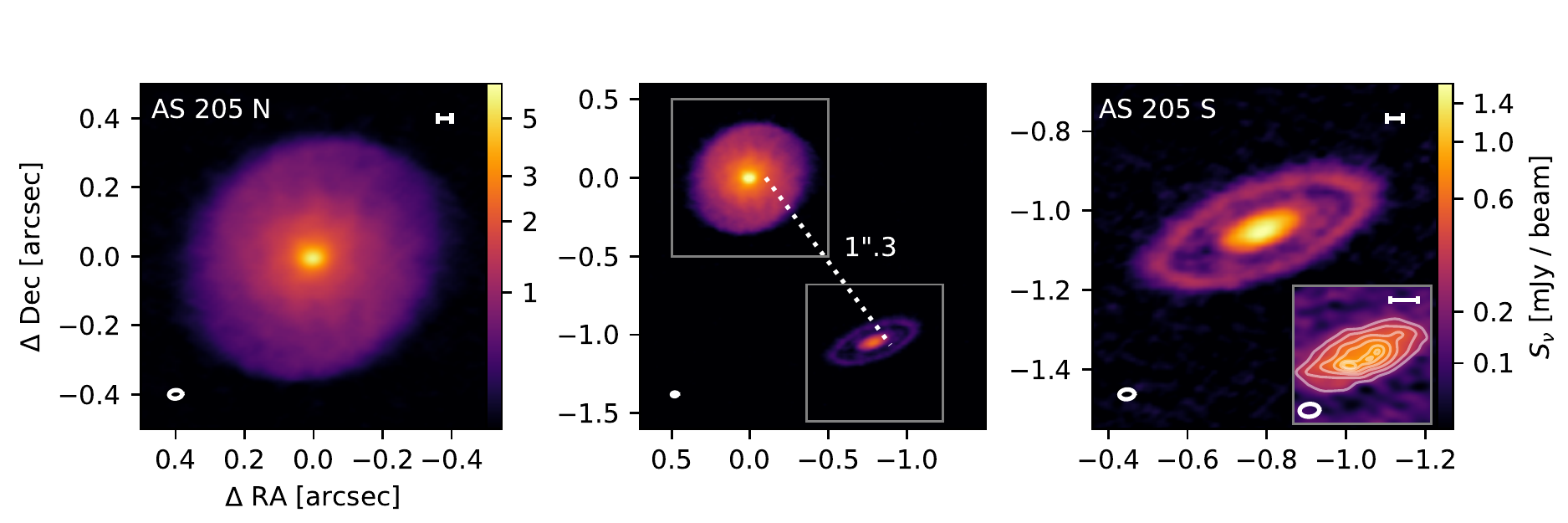}
\caption{Continuum brightness distribution in AS\,205 system. The central panel indicates the separation between the two disks on the sky, while the left/right panels show AS\,205\,N/AS\,205\,S, respectively. An inset on the lower-right corner of the AS\,205\,S panel presents a zoom to its inner disk, with observations  imaged with different imaging parameters that provide higher spatial resolution (see text).
The coordinates have their origin in the continuum peak of AS\,205\,N. The beam size is $37\times24\,$mas ($4.7\times 3.0\,$au), as shown in the lower left corner of each panel, except for the inset in the AS\,205\,S panel, with a beam size of $29\times16\,$mas ($3.7\times 2.0\,$au). In this inset, the contour levels correspond to 10, 15, 20, 25, 30 and 32$\sigma$, where $\sigma$ is the image RMS. A scale of $5\,$au is indicated by the horizontal bar in the upper right corner on the AS\,205\,N and S panels.
An arcsinh stretch is used for the color scale. 
\label{fig:AS205_mainfig} }
\end{figure*}

AS\,205 is a multiple stellar system located at a distance of $127 \pm 2\,$pc \citep{Gaia2018} in the $\rho$-Ophiuchi star-forming region. The northern and southern components (from now on, AS\,205\,N and AS\,205\,S) have been detected at a projected separation of $1\farcs{3}$ with near infrared imaging \citep[e.g.,][]{Ghez1993,Reipurth1993,McCabe2006}, and imaged in $1.3\,$mm continuum and different CO molecular lines observations \citep{Andrews2007, Salyk2014}.
The brightest of these sources at millimeter wavelengths is AS\,205\,N, a K5 pre-main-sequence star of about $0.5\,$Myr of age, with a mass of $0.87^{+0.15}_{-0.1}\, M_\odot$ \citep{Eisner2005,Andrews2018} and a mass accretion rate of 
$4 \times 10^{-8} M_\odot$ yr$^{-1}$ 
\citep{Andrews2009, Eisner2015}. It shows multiple molecular emission lines in radio and mid-infrared wavelengths \citep[e.g.,][]{Oberg2011, Salyk2014}, including water vapor lines \citep[e.g.,][]{Salyk2008, Pontoppidan2010} and organics \citep{Mandell2012}.
AS\,205\,S is itself a spectroscopic binary, with K7 and M0 spectral types and masses of $0.74$ and $0.54\,M_\odot$ \citep{Eisner2005}. 

Strong departure from Keplerian motion is detected in different molecular lines in this system and an extended emission is found around the disks, that is unlikely to arise from envelope emission, nor from a large reservoir of mass that is being accreted by these disks \citep{Salyk2014}. Instead, it might be due to a combination of disks winds and perturbations produced by the binary interaction. Given that the synthesized beam could barely separate the N and S components (beam size $\approx 0\farcs{7}$), only the large scale features of this system could be identified.

The distances to each source in the AS\,205 system used here were calculated from Gaia DR2 \citep{Gaia2018}. However, we found a difference of almost $30\,$pc between AS\,205\,N and AS\,205\,S from their Gaia parallaxes ($7.817\pm0.098$ and $6.376\pm0.185\,$mas, respectively). As it will be shown in Sect.~\ref{sec:results}, we are able to resolve the gas flow between the N and S components, previously detected in \cite{Salyk2014}, and therefore conclude that the distance between disks must allow such an interaction.
Since AS\,205\,S is an unresolved spectroscopic binary, Gaia DR2 did not account for the binary motion when calculating its parallax, which is calculated from the photocentre of each detected source \citep{Lindegren2018}.
Because of this, in the following we consider the distance to AS\,205\,N as being the same for both northern and southern sources.

\subsection{HT Lup} \label{sec:htlup}

HT\,Lup is a triple stellar system located at a distance of $154 \pm 2\,$pc \citep{Gaia2018} in the Lupus star forming region, with an age of $\approx 0.8\,$Myr \citep{Andrews2018}. Its three components, hereafter referred to as HT\,Lup\,A, B, and C, have been identified  through near infrared imaging \citep{Ghez1997,Correia2006} and interferometry \citep{Anthonioz2015}, with separations of $\sim0.1''$ between A and B, and $\sim3''$ between AB and C. Both B and C companions have lower luminosities than the primary, estimated to be $15\%$ and $9.5\%$ of that of HT\,Lup\,A \citep{Anthonioz2015}.

An extended nebulosity that resemble an arc-like structure is observed in the far infrared with Herschel photometry \citep{Cieza2013, Bustamante2015}, while cloud contamination is also found in optical spectra \citep{Herczeg2014}.
At millimeter wavelengths, the HT\,Lup system has been observed by ALMA in continuum at $890\,\mu$m and $1.3\,$mm, and in CO molecular lines \citep{Tazzari2017, Ansdell2018}. All previous observations reach an angular resolution on the order of $0\farcs{1}$, unable to resolve the closest companion, nor the individual disk structure and gas dynamics.

We estimate the distances for HT\,Lup\,A and C from Gaia DR2 \citep{Gaia2018}, with the distance to C ($154\pm3\,$pc) being consistent with the A component at the $1\sigma$ level, showing that their proximity in the sky is not a projection effect.


\section{Observations} \label{sec:data}

\begin{figure*}[ht!]
\includegraphics[width=\linewidth]{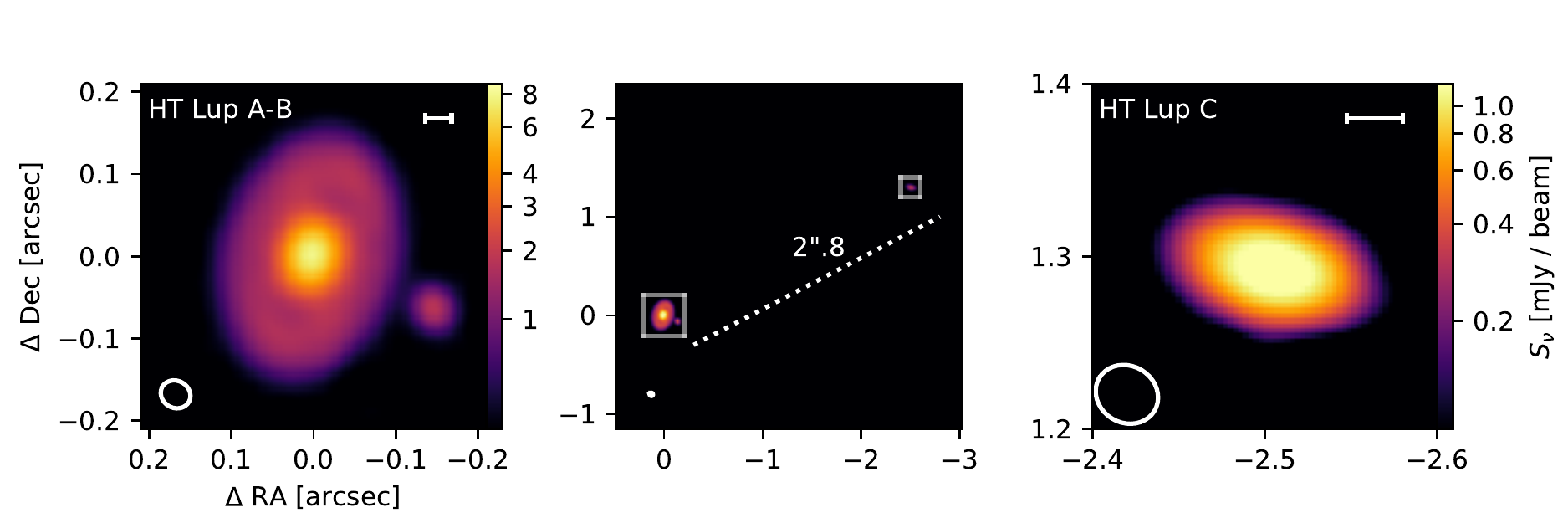}
\caption{Continuum brightness distribution in the HT\,Lup system. The central panel shows the separation between the three disks on the sky, while the left/right panels show HT\,Lup\,A-B/HT\,Lup\,C, respectively. The coordinates have their origin in the continuum peak flux of HT\,Lup\,A. The beam size is $32\times37\,$mas, as shown in the lower left of each panel, and a scale of $5\,$au is indicated by the horizontal bar in the upper right corner. An arcsinh stretch is used for the color scale. \label{fig:HTLup_mainfig}}
\end{figure*}

The datasets presented here are part of DSHARP \citep{Andrews2018}.
For AS\,205, we also include band\,6 archival data from ALMA Cycle 0 (Project number \texttt{2011.0.00531.S}), that were presented in \citet{Salyk2014}, where the CO line was also observed. A detailed description of the data acquisition and calibration can be found in \citet{Andrews2018}.

An identification of the peak position of HT\,Lup\,A was required in order to align astrometrically the different observations, following the procedure in \citet{Andrews2018}. However, HT\,Lup\,A and B components can only be resolved using the long-baselines datasets, therefore, the peak location of \HT could not be identified in the shortest baselines datasets. Given that HT\,Lup\,C was $2\farcs{8}$ apart and isolated, we used its position as an alignment reference, before starting the self-calibration process.
The final images are centered at (J2000) RA\,=\,16h\,11m\,31.352s, Dec = $-18$d\,38m\,26.233s for AS\,205, and for HT\,Lup the center is at RA\,=\,15h\,45m\,12.847s, Dec\,=\,$-34$d\,17m\,31.01s.

After self-calibration, we generated continuum images using the \texttt{tclean} task in CASA\,5.1 \citep{mcmullin07}. For AS\,205 we chose a robust parameter of $0.5$, resulting in the $1.25\,$mm continuum images displayed in Figure~\ref{fig:AS205_mainfig}, with a beam size of $37\times 24\,$mas ($4.7\times 3.0\,$au). In addition, we also created an image with uvtaper of $40\times0\,$mas and PA=$90^\circ$ (measured from north to east) to circularize the beam and increase the signal-to-noise ratio (SNR) in extended emission, which was only used to study the substructure of AS\,205\,N, while an image with robust $= -0.5$ was used to take a look into the compact central emission of AS\,205\,S. The first continuum image has an SNR of 390 with an rms of $16\,\mu$Jy/beam, the second has an SNR of 645 with an rms of $18\,\mu$Jy/beam, and the third has an SNR of 136 and an rms of $26\,\mu$Jy/beam.

While most of the CO maps in our survey are imaged with beams of $\approx 100\,$mas, for the disks around these multiple systems we synthesized beams with smaller size, in order to distinguish structures inside the most compact sources. Channel maps from CO in AS\,205 were generated using a robust parameter of $1.$ to obtain a beam of $90\times67\,$mas ($\approx 8.5\,$au at best), which led to a rms of $1.42\,$mJy/beam per velocity channel, and a peak SNR of $35.3$. The AS\,205 channel maps are presented in Figure~\ref{fig:AS205_CO_gallery} of the appendix.

For HT\,Lup continuum we chose a robust parameter of $0.5$, and the resulting 1.3\,mm images are displayed in Figure~\ref{fig:HTLup_mainfig}, with a beam size of $37\times32\,$mas ($5.7\times 4.9\,$au), a rms of 14.1\,$\mu$Jy/beam and a peak SNR of $585$.
In order to obtain the cleanest possible CO map, we excluded baselines smaller than $150\,$m, thus, decreasing the sensitivity to scales larger than $2\arcsec$, emphasizing compact emission.
The CO images used a robust parameter of 1.5, and we additionally applied an uv-tapering of $20\times5\,$mas with $PA=150^\circ$, resulting in a beam of $53\times50\,$mas ($\approx 8\,$au). 
The rms of this spectral cube is $1.2\,$mJy/beam per velocity channel, with a peak SNR of $10.5$. The HT\,Lup channel maps are presented in Figure~\ref{fig:HTLup_CO_gallery} of the appendix.


\section{Results} \label{sec:results}

\begin{table*}
  \begin{center}
    \caption{Results of 2D Gaussian fit to each continuum disk.}
    \begin{tabular}{c|c|c|c|c|c|c}\toprule
	 Source & RA (J2000) & Dec (J2000) & major axis\tablenotemark{b} & minor axis\tablenotemark{b} & $i$ & $PA$ \\
	 \midrule
	 AS\,205\,N\tablenotemark{a} & 16:11:31.352 & -18:38:26.34 & $0\farcs{414}\pm0.006$ & $0\farcs{388}\pm0.006$ & $20.1^\circ\pm3.3^\circ$ & $114.0^\circ\pm11.8^\circ$ \\
     AS\,205\,S & 16:11:31.296   & -18:38:27.29 & $0\farcs{185}\pm0.006$ & $0\farcs{077}\pm0.003$ & $66.3^\circ\pm1.7^\circ$ & $109.6^\circ\pm1.8^\circ$\\
     \midrule
     HT\,Lup\,A & 15:45:12.847   & -34:17:31.01 & $0\farcs{156}\pm0.010$ & $0\farcs{104}\pm0.007$ & $48.1^\circ\pm4.5^\circ$ & $166.1^\circ\pm6^\circ$ \\
     HT\,Lup\,B & 15:45:12.835   & -34:17:31.08 & $0\farcs{032}\pm0.001$ & $0\farcs{022}\pm0.002$ & $44.9^\circ\pm4.9^\circ$ & $8.3^\circ\pm7.5^\circ$  \\
     HT\,Lup\,C & 15:45:12.645   & -34:17:29.72 & $0\farcs{059}\pm0.001$ & $0\farcs{025}\pm0.001$ & $65.5^\circ\pm0.9^\circ$ & $78.8^\circ\pm0.8^\circ$  \\
	\bottomrule
    \end{tabular}
  \end{center}
\label{tab:gauss_fit}
\tablenotetext{a}{For this target the fit was done on the short-baseline images only, to avoid including substructure in the 2D Gaussian fit.}
\tablenotetext{b}{Deconvolved gaussian values.}
\end{table*}

\subsection{Continuum emission in the AS\,205 system}

In the continuum emission, the AS\,205 system resolves into two disks whose peaks are separated by $1\farcs{313}$, or $168\,$au in projected separation (central panel, Figure~\ref{fig:AS205_mainfig}), along a position angle of $217^\circ$. The AS\,205\,N disk is not azimuthally symmetric, instead, a spiral-like pattern with low contrast is observed (left panel, Figure \ref{fig:AS205_mainfig}). The AS\,205\,S disk is fainter and smaller in angular size than AS\,205\,N, and it exhibits a narrow ring around an inner disk (right panel, Figure \ref{fig:AS205_mainfig}), where a cavity is observed when imaged at high angular resolution (see inset on the right panel, Figure \ref{fig:AS205_mainfig}).

In other DSHARP targets with spiral features \citep[e.g.\ Elias 2-27 or IM\,Lup,][]{Huang2018b} there are symmetric substructures (bright or dark rings) that can be used to constrain the geometry of the disk. However, the lack of symmetric features in AS\,205\,N implies that we have to use a different method to constrain the disk inclination ($i$) and position angle ($PA$).
Given that the two sources are well separated in the sky, we fitted a 2D Gaussian model to each disk using the CASA task \texttt{imfit}. For AS205\,N, we used the continuum image generated while excluding the longest baseline dataset, with a beamsize of $270\times227\,$mas, to avoid including any asymmetric feature. The best-fit values are given in Table \ref{tab:gauss_fit}. With the values of $i$ and $PA$ derived, we find that the angular momentum vectors of the disks are misaligned by either $46^\circ$ or $94^\circ$ \citep[see equation in e.g.,][]{Jensen2014}, depending on whether the two disks share the same near side (the disk side closer to the observer).

\subsubsection{Spirals in AS\,205 North}
\label{sec:as205n}

\begin{figure}[t!]
\includegraphics[width=\linewidth]{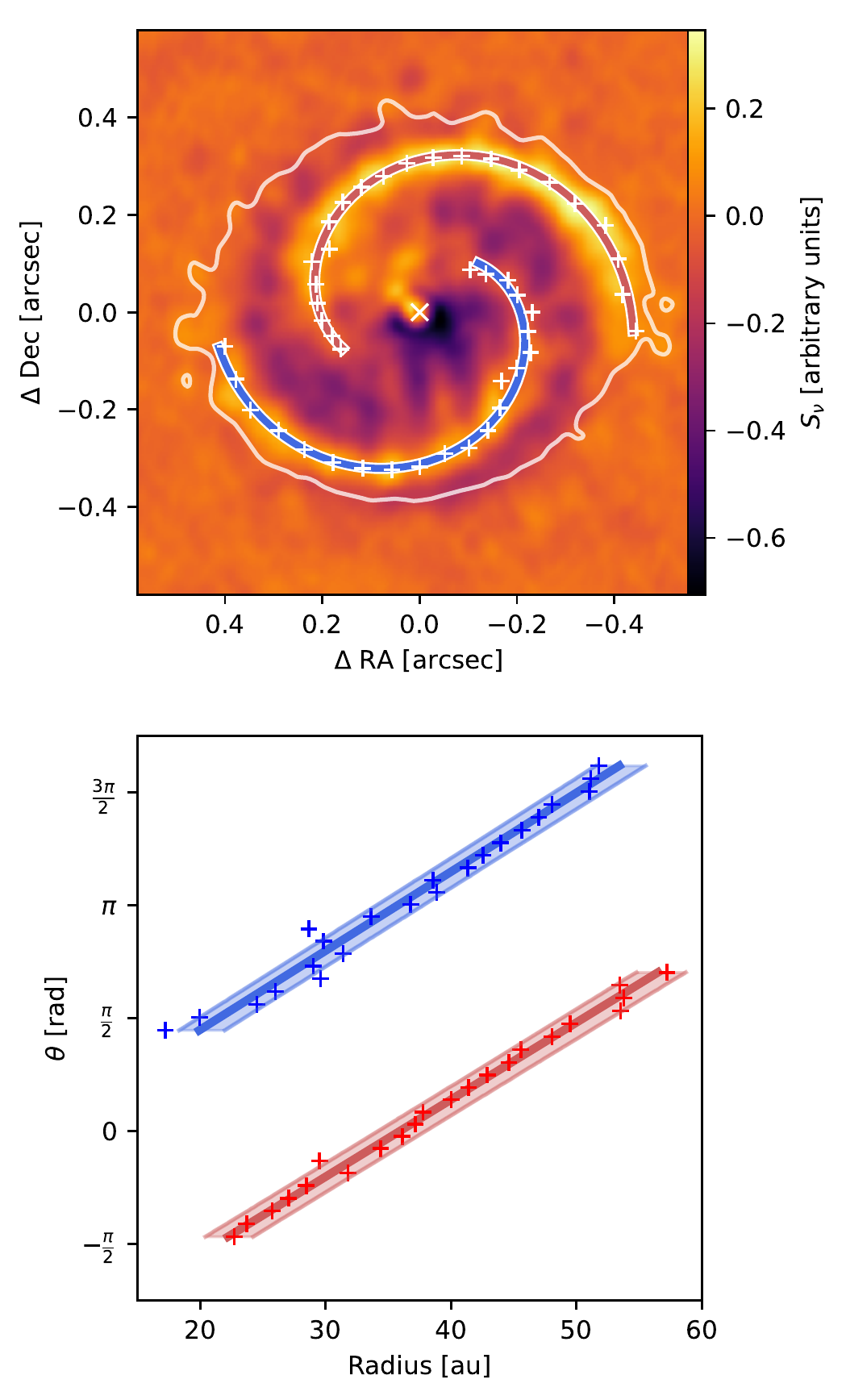}
\caption{Top panel: Continuum emission of AS\,205\,N after subtraction of its azimuthally averaged radial profile. The plus symbols mark the location of the maximum of emission along the spiral arms, with the best-fit archimedean model in red (NW) and blue (SE) lines. The contour level surrounding the disk marks the 5$\sigma$ level of the original image (before subtraction of the radial profile). The center of the spirals was fixed, and its position is marked with a white cross. 
Bottom panel: Deprojected spiral wake, with colors as in the top panel. The best fit model is shown with a solid line, and the shaded region represents the $1\sigma$ uncertainties of our best-fit.
\label{fig:AS205_spirals_bestfit}}
\end{figure}

We calculate the azimuthally averaged radial profile of the continuum emission, considering $i$ and $PA$ as constrained above, and using the peak of emission as center (which coincides with the peak of the 2D Gaussian fitted within $3\,$mas, about 1/10 of the beam size).
After subtracting this radial profile from the continuum image, a clear spiral structure is revealed, as shown in Figure~\ref{fig:AS205_spirals_bestfit}. These spiral features can be traced between $\sim20$ to $55\,$au, beyond this radius the intensity of the continuum disk goes below the $3\sigma$ level.
To trace the north-west (NW) and south-east (SE) spiral arms, we define a set of discrete points that correspond to the peak of emission along the radial direction, spaced by $10^\circ$ in azimuth (which corresponds to one synthesized beam at $\sim30$\,au) and identified where the disk emission is above $3\sigma$.

To characterize each spiral, we consider models of a logarithmic spiral defined as:
\begin{equation}
r = r_0 \cdot \exp(b \theta) 
\end{equation}
and of an Archimedean spiral, defined as:
\begin{equation}
r = r_0 + b \theta
\end{equation}
\noindent where $\theta$ is the azimuthal angle, $r_0$ the radius when the angle is $0$, and $b$ relates to the pitch angle $\mu$ of the spiral. 
For the logarithmic spiral, the pitch angle is constant along all radii and it is calculated as $\mu = \arctan{(1/b)}$, while for the Archimedean model, the pitch angle depends on the radius as $\mu = b / r$.

The NW and SW spirals are assumed to share the same center (located at the peak of emission), while the spiral parameters $r_0$ and $b$ are fitted separately for each arm to test if these are symmetric or not. We also include the observed inclination ($i$) and position angle ($PA$) as free parameters, assuming both spirals share the same geometry. Therefore, we have 6 free parameters ($r_{0,NW}$, $r_{0,SE}$, $b_{NW}$, $b_{SE}$, $i$, $PA$). 
To fit the spiral prescriptions above, we use an MCMC routine based on \texttt{emcee} \citep{emcee2013}. A flat prior probability was used for all parameters. For each fit, we use 250 walkers with two consecutive burning stages of 1000 and 500 steps, and then 1500 steps to sample the parameter space. The results of the logarithmic and Archimedean spiral model fit are given in Table \ref{tab:AS205_spiral_fit}. We note that the reduced $\chi^2$ in the Archimedean spiral model is factor of $1.5$ better than in the logarithmic model. Figure \ref{fig:AS205_spirals_bestfit} shows the best-fit Archimedean spiral and the location of emission maxima in the image (top panel) and in polar coordinates (bottom panel).

We note that we tested a model that allows for an offset of the spirals with respect to the center (two additional free parameters). The model finds an offset that is smaller than $\approx 3\,$au, with the NW and SE spirals pitch angle 
differing but consistent with each other within 1$\sigma$. Since the reduced $\chi^2$ is comparable to the model with fixed center, we chose to use the latter for simplicity.

\begin{table}
  \begin{center}
    \caption{Best fit and 1$\sigma$ uncertainties from the fit of the spiral shape in AS\,205\,N. The pitch angles $\mu$ are calculated from $b$. For the Archimedean model, $\mu$  is calculated at 35\,au.} 
    \begin{tabular}{c|c|c}\toprule
	  Parameter  & Log. & Arch. \\
	  \midrule
      $r_{0,NW}$ & $26.1_{-1.1}^{+1.3}\,$au  & $36.0_{-1.8}^{+2.0}\,$au \\
      $r_{0,SE}$ & $11.9_{-0.6}^{+0.7}\,$au  & $7.2_{-1.6}^{+2.1}\,$au \\
      $b_{NW}$   & $0.244_{-0.006}^{+0.005}$ & $9.32_{-0.15}^{+0.23}$ \\
      $b_{SE}$   & $0.246_{-0.006}^{+0.005}$ & $9.10_{-0.23}^{+0.16}$ \\
      $\mu_{NW}$ & ${13.9^\circ}_{-0.3}^{+0.3}$ & $15.3^\circ$ at 35\,au \\
      $\mu_{SE}$ & ${13.7^\circ}_{-0.3}^{+0.3}$ & $14.9^\circ$ at 35\,au \\
      $i$ & ${15.1^\circ}_{-3.2}^{+1.9}$        & ${14.3^\circ}_{-5.3}^{+1.3}$ \\
      $PA$ & ${-9.1^\circ}_{-8.7}^{+10.7}$      & ${59.6^\circ}_{-10.3}^{+13.0}$ \\
	\bottomrule
    \end{tabular}
  \end{center}
\label{tab:AS205_spiral_fit}
\end{table}

We calculate the contrast between the spiral and inter-spiral region, by comparing the intensity of each arm with the lower 5\% intensity of a ring at the same radial distance. We found the arms to be of low contrast, with only a factor of $1.4$ and $1.3$ (median value) between the spiral and the inter-spiral region, for the NW and SE spirals, respectively. The contrast between the NW and SE spirals is small, $\approx 1.1$ on average.

\subsubsection{A ring in AS\,205 South} \label{subsec:AS205S}

The disk around AS\,205\,S is also well resolved in the continuum. The peak of emission for this component is $2.4\,$mJy/beam ($185\sigma$), which is only $29\%$ of the AS\,205\,N peak, the mean surface brightness of the ring is around $57.6\,\mu$Jy/beam ($32\sigma$).

\begin{figure}[t!]
\includegraphics[width=\linewidth]{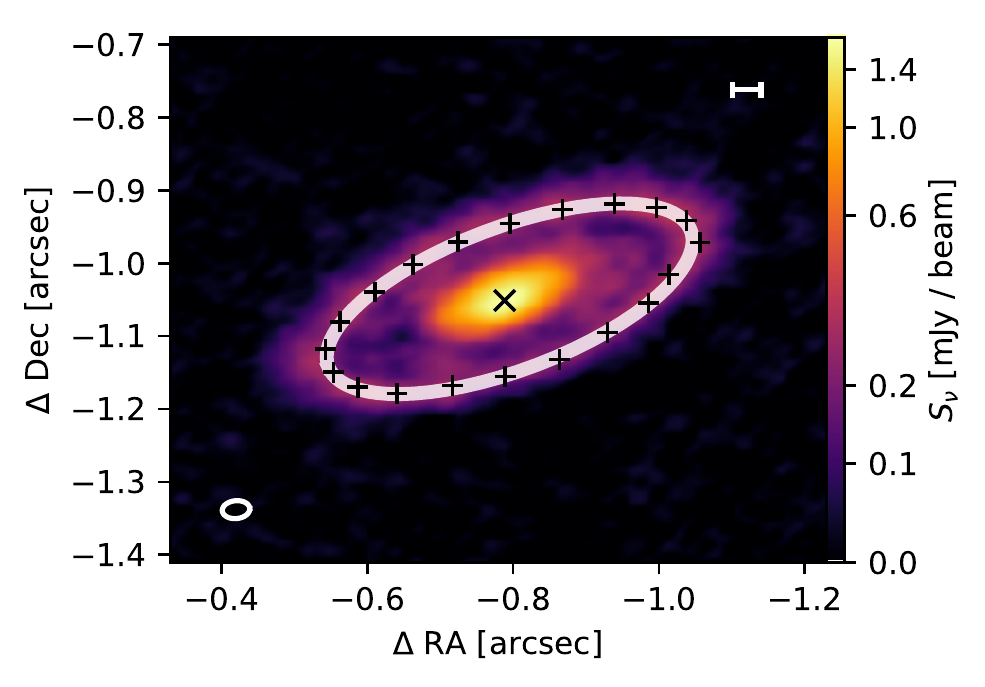}
\caption{Fit of an ellipse to the AS\,205\,S ring in the outer disk. The plus symbols mark the maximum emission along the ring, while the white line shows the best fit. At the center, the black cross marks the best-fit center of the ellipse. The beam is given in the bottom left corner, and a $5\,$au scale bar is in the upper right corner for comparison.
\label{fig:AS205_ring_fit}}
\end{figure}

At equally spaced intervals of $18^\circ$, i.e. points are roughly spaced by one synthesized beam, 
we search for the position of the maximum emission along the ring. The points were then fitted with an ellipse using a MCMC routine based on \texttt{emcee} \citep{emcee2013}, letting the center, major axis, minor axis, and position angle free to vary. The walkers and steps used are similar to the spiral fit. 
The  points selected along the ring and the best-fit model are shown in Figure~\ref{fig:AS205_ring_fit}, while the best-fit parameters are given in Table~\ref{tab:AS205_ring_fit}.

\begin{table}
  \begin{center}
    \caption{Results from the MCMC search for best-fit parameters of a ring in AS\,205\,S.  Errors correspond to $1\sigma$. Note that the major axis corresponds to the radius of the ring.}
    \begin{tabular}{c|c}\toprule
	  Parameter & Value.  \\
	  \midrule
      $\Delta x$ & ${-0.42}_{-0.3}^{+0.22}$au \\
	  $\Delta y$ & ${0.23}_{-0.18}^{+0.17}$au \\
      major axis & ${33.8}_{-0.4}^{+0.3}$    \\
      minor axis & $12.4 \pm 0.2\,$au    \\
	  $i$ & ${68.4^\circ}_{-0.7}^{+0.5}$ \\
	  $PA$  & ${110.6^\circ}_{-0.4}^{+0.5}$ \\
	\bottomrule
    \end{tabular}
  \end{center}
\label{tab:AS205_ring_fit}
\end{table}

From the fit, we find that the ring center matches the peak flux within $3\,$mas ($\sim1/10$ of the beam size), and that the ring inclination and position angle are also in good agreement with the  values obtained with the 2D Gaussian fit to the image.

\subsection{Gas emission in the AS\,205 system}

\begin{figure}[t!]
\includegraphics[width=\linewidth]{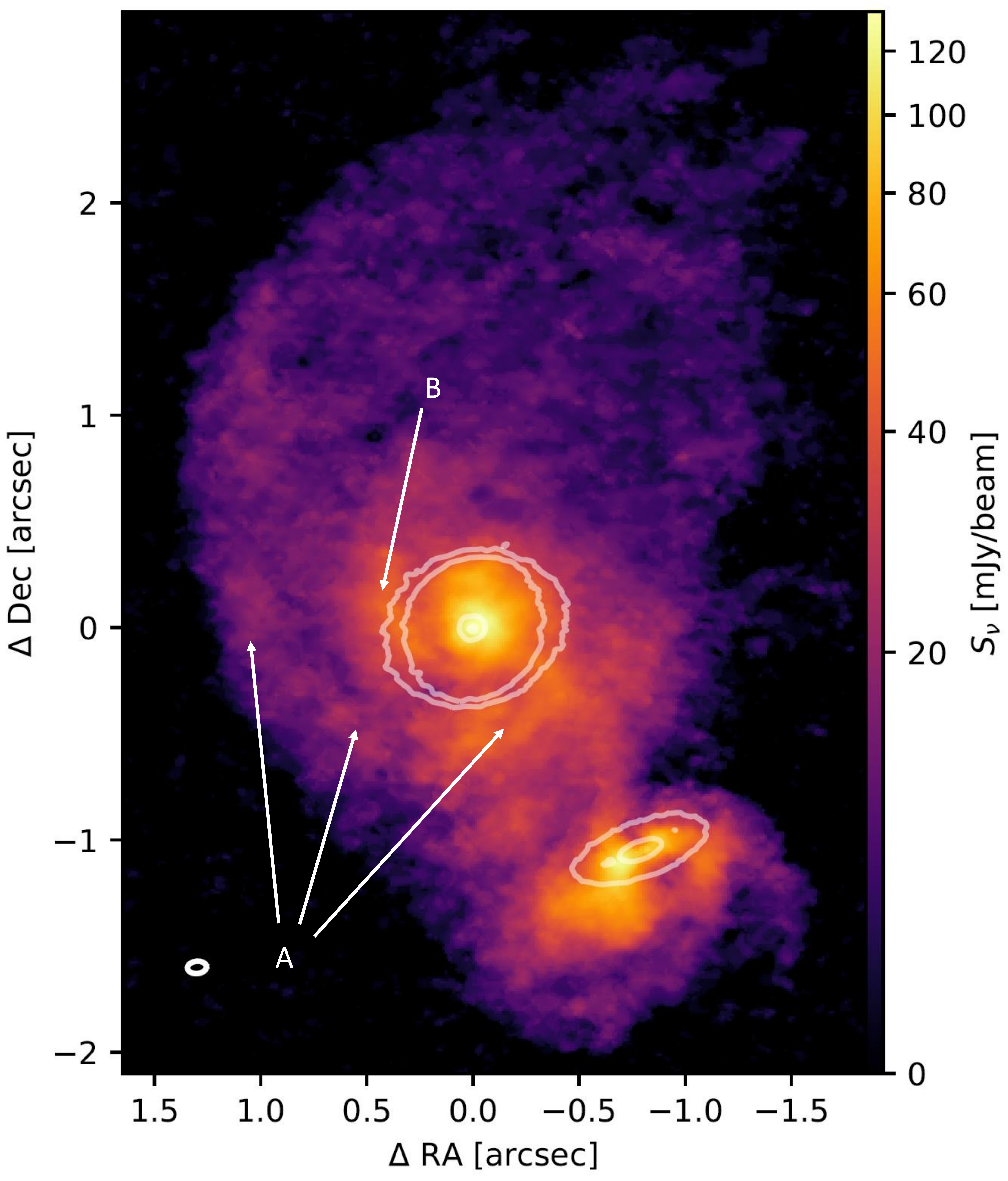}
\caption{Integrated emission map (moment zero) of the CO spectral cube in the AS\,205 system. The two main arcs of emission in AS\,205\,N are labeled A and B (See Figure \ref{fig:AS205_CO_gallery}). The contour levels represent $5$, $25$ and $300\sigma$ of the continuum emission, for comparison.
\label{fig:AS205_mom0}}
\end{figure}

\begin{figure*}[ht!]
\includegraphics[width=\linewidth]{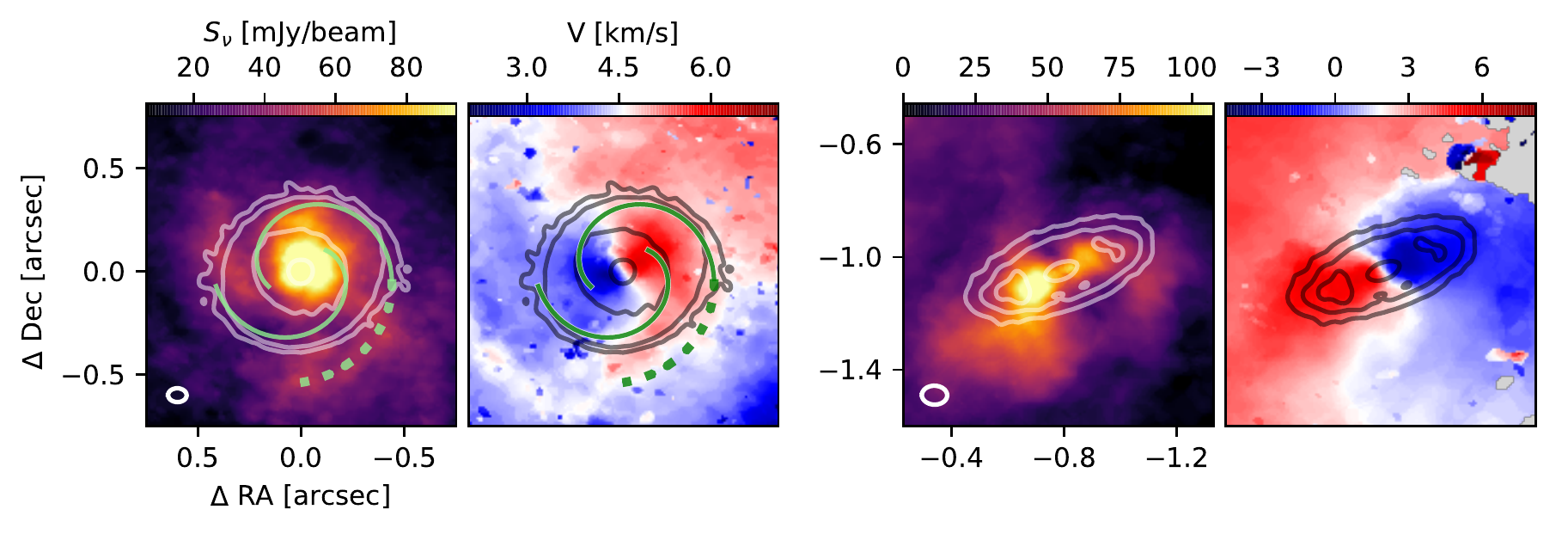}
\caption{Moment 0 and moment 1 images of the CO emission in AS\,205\,N (left panels) and AS\,205\,S (right panels). The beam size ($57\times54$ mas) is shown at the bottom-left corner of each moment 0 image. For AS\,205\,N, we draw in green the best archimedean fit to continuum spirals, and the NW spiral is extended as a dashed line for comparison.
The contour lines mark the $5\sigma$, $25\sigma$, $120\sigma$ and $300\sigma$ level in the continuum image.
\label{fig:AS205_moms}}
\end{figure*}

We detect CO emission from this system from $v=-7.6\,$km/s to $+12.0\,$km/s, with little cloud contamination over this velocity range. Figure~\ref{fig:AS205_mom0} shows the integrated intensity (moment 0) while Figure~\ref{fig:AS205_moms} also presents the intensity-weighted velocity field (moment 1) computed from the CO datacube, clipping at $3\sigma$ and including only channels with detected emission (see Figure~5.10 of \citet{Andrews2018} for all channel maps,and Figure~\ref{fig:AS205_CO_gallery} in appendix for channels of interest).
Evidence of tidal interaction is clearly seen in the gas tracer, with CO emission between the two continuum sources on channels between $v=3.25\,$km/s and $5.35\,$km/s.
The AS\,205\,N disk shows a butterfly pattern characteristic of Keplerian motion around the central star, which allow us to estimate its systemic velocity to be $\approx 4.5\,$km/s (emission above $3\sigma$ is observed from $v=-0.1\,$km/s to $+8.5\,$km/s). However, this Keplerian pattern only holds inside the region where the continuum emission is above $3\sigma$, at approximately $60\,$au from disk center. Outside this region, we observe extended emission and several arc-like structures that extend to the outskirts of the disk (at most at $410\,$au, $\approx 3\farcs{2}$).

In the north and south sides of the continuum disk, there are  arc-like structures in CO that resemble spiral arms. The most prominent arc (that starts in the west and turns clock-wise to the south and then east, labeled A in Figure~\ref{fig:AS205_mom0}), roughly coincides with the NW continuum spiral, as is shown with the best-fit model for this spiral in dotted lines in Figure~\ref{fig:AS205_moms} (left panels). However, further out than $\sim80$au ($\sim 0\farcs{5}$) the arc does not have the same opening angle as the NW continuum spiral.
In the moment 1 map from Figure~\ref{fig:AS205_moms}, the trace of arc A appears to have constant velocity ($\sim 4.5\,$km/s) over its extension outside of the continuum disk. In the moment 0 map, another spiral-like structure can be distinguished towards the east (labeled B in Figure~\ref{fig:AS205_mom0}), but this feature is not co-located with the best-fit SE continuum spiral. Furthermore, arc B is not clearly observed across channels maps, and no velocity structure that corresponds to this arc can be distinguished in the moment 1 map either.

As can be seen in the moment 1 map of the CO in AS\,205\,S (Figure~\ref{fig:AS205_moms}, rightmost panel), the southern component shows disk rotation, but quite perturbed. First, due to its high inclination ($\sin i \approx 0.9$) the inner disk emission can be seen at high velocities from $v=-7.6\,$km/s to $+12.0\,$km/s, which is about a factor of two wider velocity range than for AS\,205\,N. Non-Keplerian motion is seen in the south-east of AS\,205\,S over all channels. At velocity channels near 4.3\,km/s, the gas emission appears as a broad arc towards the south, better appreciated as the bright emission in the south of the AS\,205\,S moment 0 map.

\subsection{Continuum emission in the HT\,Lup system}

Three components are detected in this system, Figure~\ref{fig:HTLup_mainfig} shows the 1.3\,mm continuum map where we are able to spatially resolve the dust continuum emission around the closest pair:  HT\,Lup\,A and B. The angular separations between HT\,Lup\,A and B, and HT\,Lup\,A and C are $0\farcs{161}$ and $2\farcs{82}$, respectively, which corresponds to a projected separations of $25$ and $434\,$au.
We fitted a 2D Gaussian using \texttt{imfit} in CASA to derive inclinations and position angles for all disks, listed in Table \ref{tab:gauss_fit}. From these values, and following the same procedure used in AS\,205, we estimate the misalignment between the angular momentum of the disks to be either $91^\circ$ or $164^\circ$ for \HT and B, and $76^\circ$ or $108^\circ$ for A and C.

\subsubsection{Spirals in HT\,Lup\,A}

\begin{figure}[tb]
\includegraphics[width=\linewidth]{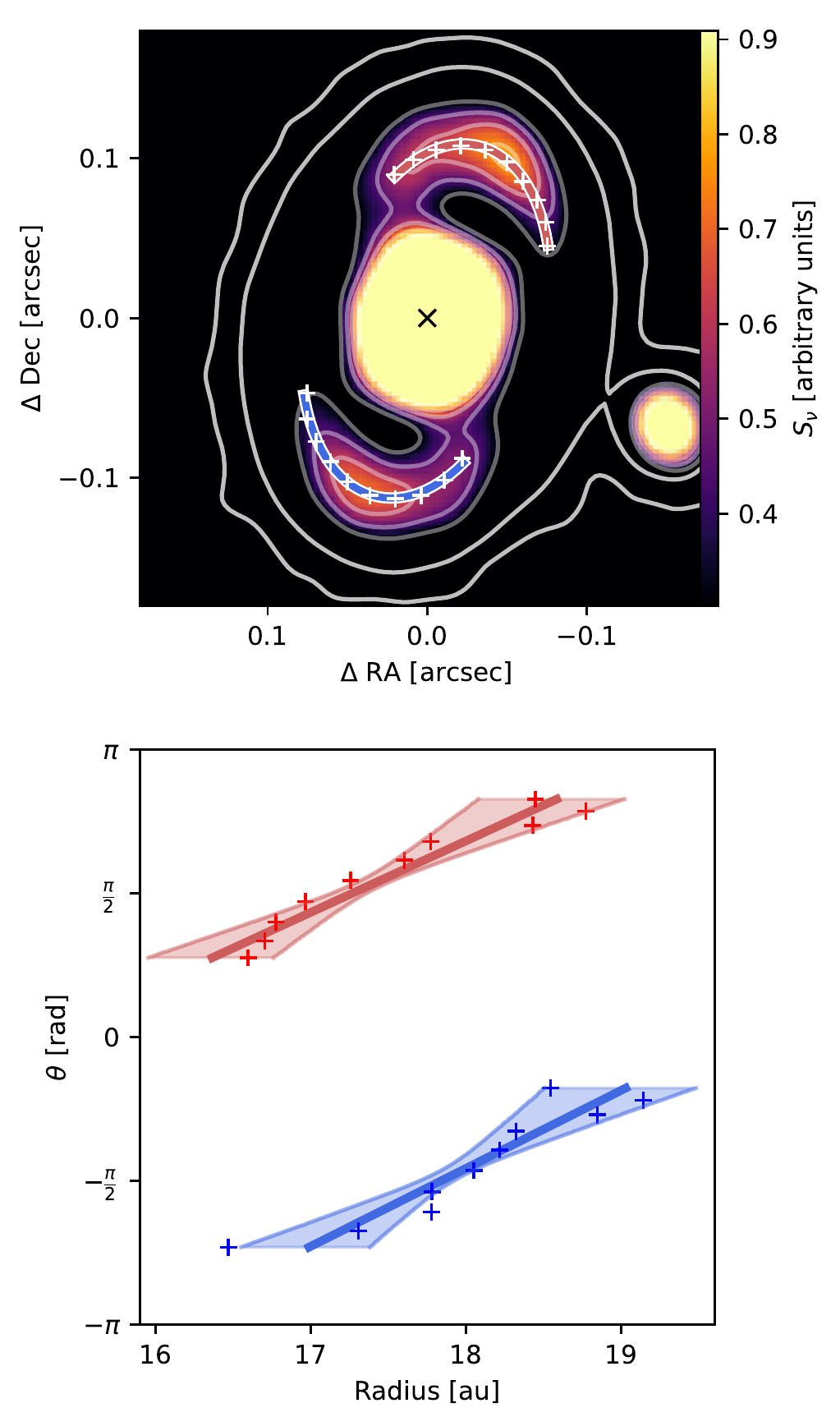}
\caption{Top panel: The HT\,Lup\,A/HT\,Lup\,B unsharp-masked continuum image. The plus symbols mark the maximum emission along the spirals. In colored solid line the best Archimedean spiral models. Solid outer lines show contour levels at $5\sigma$ and $28\sigma$ of the original continuum image. The color scale was chosen to emphasize spiral arms.
Bottom panel: Deprojected spiral wake, with colors as in the top panel. The best fit model is shown as a solid line, and the shaded region represents the $1\sigma$ uncertainties of our best-fit. 
\label{fig:HTLup_spirals_bestfit} }
\end{figure}

HT\,Lup\,A is the brightest and more extended disk in the system, with emission above $3\sigma$ detected up to $33\,$au ($\approx0\farcs{21}$) from the center.
Computing an average radial profile of emission on this disk is difficult due to the presence of the companion at close distance (HT\,Lup\,B). Thus, to subtract the overall disk emission and enhance the non axisymmetric features in the disk, we use an unsharped masking technique. We first convolve the image with a Gaussian of $66\,$mas FWHM and then subtract it from the original continuum image, multiplying by a weighting factor of 0.95. We chose these unsharp masking parameters as 
they better enhanced the low-contrast spiral features (convolution with larger Gaussians smooths out the disk emission excessively, smaller Gaussians do not smooth out the non-symmetric features and these end up being subtracted instead of enhanced).
The resulting image, shown in Figure~\ref{fig:HTLup_spirals_bestfit}, reveals an underlying spiral structure.
We trace the arms as in AS\,205\,N, finding the maxima along radial directions separated by $8^\circ$. The spirals extend from $\approx16.5$ to $\approx19\,$au in radius, and each arm covers an azimuthal extent of $\approx100^\circ$.

Following the procedure in Sect.~\ref{sec:as205n}, we fit a logarithmic and Archimedean spiral model, with the best-fit parameters presented in Table~\ref{tab:HTLup_spiral_fit}. The results show spirals with low pitch angles and quite symmetric, however, they are so compact that possible asymmetries might remain unresolved by our observations. The inclination and position angle are in agreement with the values obtained from Gaussian fitting.

\begin{table}
  \begin{center}
    \caption{Best fit and 1$\sigma$ uncertainties from the fit of the spiral shape in HT\,Lup\,A. The pitch angles $\mu$ are calculated from $b$. For the Archimedean model, $\mu$  is calculated at 18\,au.}
    \begin{tabular}{c|c|c}\toprule
	  Parameter  & Log. & Arch. \\
	  \midrule
      $r_{0,N}$ & $15.4_{-1.8}^{+1.9}\,$au & $15.3 \pm 0.8\,$au \\
	  $r_{0,S}$ & $19.7_{-0.7}^{+0.8}\,$au & $19.7_{-0.8}^{+0.7}\,$au \\
	  $b_{N}$  & $0.073_{-0.03}^{+0.025}$ & $1.28_{-0.51}^{+0.5}$ \\
      $b_{S}$  & $0.064_{-0.03}^{+0.025}$ & $1.17_{-0.51}^{+0.5}$ \\
      $\mu_{N}$ & ${4.15^\circ}_{-1.7}^{+1.4}$ & $4.1^\circ$ at 18\,au \\
	  $\mu_{S}$ & ${3.69^\circ}_{-1.7}^{+1.4}$ & $3.7^\circ$ at 18\,au \\
	  $i$ & ${52.2^\circ}_{-0.9}^{+0.6}$ & ${53.0^\circ}_{-0.7}^{+0.6}$ \\
	  $PA$   & ${14.3^\circ}_{-1.8}^{+2.3}$ & ${14.2^\circ}_{-2.1}^{+2.2}$ \\
	\bottomrule
    \end{tabular}
  \end{center}
\label{tab:HTLup_spiral_fit}
\end{table}

\subsubsection{Companions: HT\,Lup\,B and HT\,Lup\,C}

HT\,Lup\,B appears barely spatially resolved, its peak intensity ($1.85 $mJy/beam, $131\sigma$) is 23\% of the HT\,Lup\,A peak intensity. From the 2D Gaussian fitting we obtain a deconvolved FWHM size of $31\pm2\,$mas, corresponding to a disk size of $\sim5\,$au.

The farthest companion, HT\,Lup\,C, is the faintest source in the system, with a peak intensity of $1.6\,$mJy/beam (20\% of the peak of HT\,Lup\,A) and a total integrated flux of $3.48$\,mJy. From the 2D Gaussian fit, we measure a deconvolved size of $59\pm1\,$mas, which corresponds to a disk size of $\sim9\,$au.

\subsection{Gas emission in the HT\,Lup system}

\begin{figure}[t!]
\includegraphics[width=\linewidth]{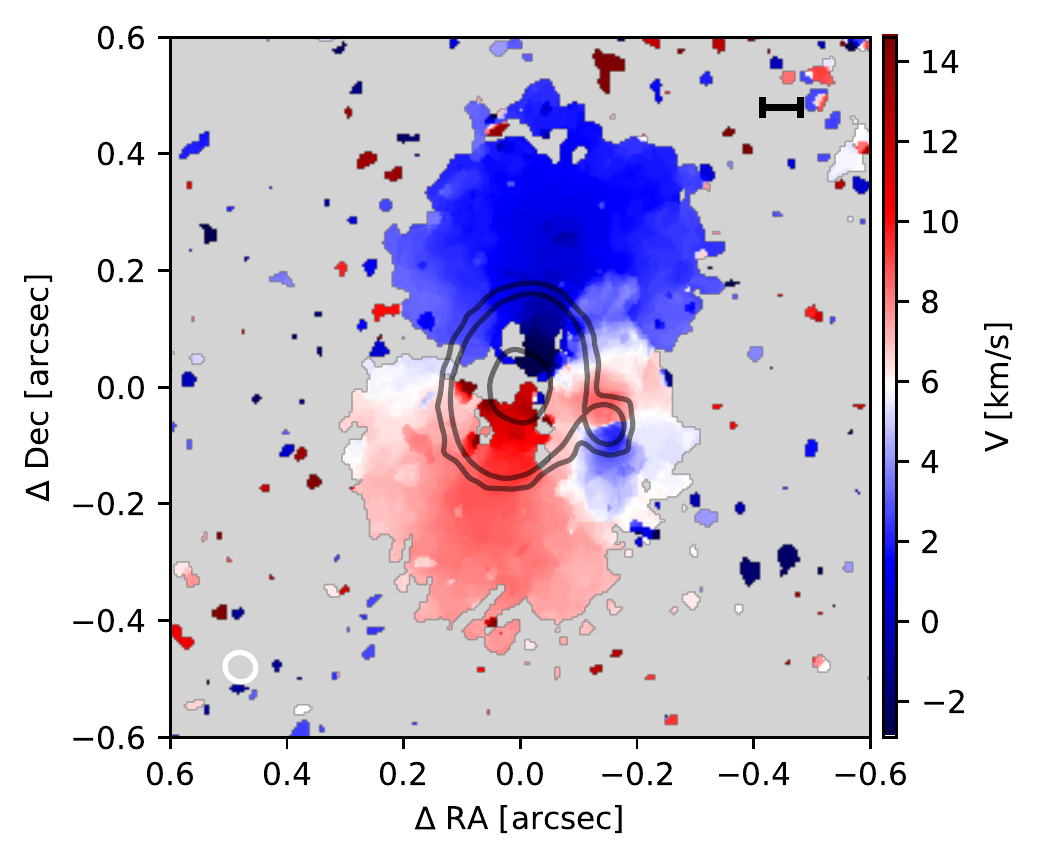}
\caption{Moment 1 of the CO emission in the region of HT\,Lup\,A and HT\,Lup\,B. The beam size is $0\farcs{053} \times 0\farcs{049}$, presented in the bottom left corner of the image, as well as a $10\,$au scale bar at the upper right corner. The contour lines mark the $5\sigma$, $28\sigma$ and $150\sigma$ level of the continuum image, for comparison. \label{fig:HTLup_mom1} }
\end{figure}

A map of the CO emission was obtained following the DSHARP procedure \citep{Andrews2018}.  However, the CO was found to be highly contaminated by extended cloud emission and foreground absorption near the systemic velocity ($v_{sys}\approx 5.5\,$km/s), at the level of completely erasing the signal from the disks between $3.75$ and $4.8\,$km/s.
(for all channel maps see Figure~5.1 of \citet{Andrews2018}, while channels of interest are in Figure~\ref{fig:AS205_CO_gallery} in the appendix).
For HT\,Lup\,A, blueshifted emission is seen in the north, while redshifted emission appears in the south. The opposite is observed for HT\,Lup\,B. This is better seen in the first moment map of CO emission, presented in Figure~\ref{fig:HTLup_mom1}, in which the disks appear to be counter-rotating.
HT\,Lup\,C is also observed in the CO, with emission detected from $v=2.7$ to $9.7\,$km/s, extending $\sim0\farcs{2}$ along its major axis, which lies horizontally as expected from its continuum shape.

\section{Discussion \label{sec:discussion}}

\subsection{Ruling out chance alignment}

Given that the AS\,205 components have different parallaxes as measured by Gaia DR2 (resulting in a difference of almost 30\,pc in distance, see Section \ref{sec:as205}), and that for the HT\,Lup system there are no constraints on the distance to source B (while the A and C components have consistent Gaia parallaxes), one could argue that the observed vicinity of these pairs is due to chance alignment. 
In the case of AS\,205, an interaction between AS\,205\,N and S is observed in CO emission in these observations, as well as in earlier works \citep{Salyk2014}.

In the case of HT\,Lup, with speckle imaging,  \cite{Ghez1997} find that the angular separation between HT\,Lup\,A and B is $0\farcs{107} \pm 0\farcs{007}$ in 1997, while \cite{Correia2006} measure a separation of $0\farcs{126} \pm 0\farcs{001}$ with data from 2004, using VLT observations. In this work, we constrain a separation of $0\farcs{161} \pm 0\farcs{003}$ with data from 2017. Thus, in the span of $\sim20$ years, HT\,Lup\,A and B have changed their separation by $\sim50$mas. Given the proper motion of HT\,Lup\,A \citep[$\mu_{R.A.} = -13.63 \pm 0.13$ mas/yr,  $\mu_{Dec} = -21.61 \pm 0.08$ mas/yr,][]{Gaia2018}, if the pair was aligned by chance then their separation should have changed by $\sim500$\,mas over this timespan, an order of magnitude larger than measured. We therefore conclude that most likely, the HT\,Lup\,A and B stars as well as the AS\,205\,N and S components are not aligned by chance.

\subsection{Substructures in the dust continuum emission}

The high resolution observations of the multiple systems HT\,Lup and AS\,205 have allowed us, for the first time, to directly constrain the type of substructures present in protoplanetary disks that have an ongoing interaction. In the following, we discuss the different substructures found in gas and dust tracers, and compare with other systems or numerical simulations.

\subsubsection{Spiral arms}

The spiral arms observed in the primary components of these systems are quite different from each other, and only \AS looks similar to the spirals observed in single systems from the DSHARP sample \citep[e.g.\ WaOph\,6 and IM\,Lup,][]{Huang2018b}.

The HT\,Lup\,A disk is quite compact ($32\,$au radius) and the radial extent of the spirals is only $\sim4$\,au (about 10\% of the disk size). However, to describe the maxima of emission in the unsharped masking image, our modeling prefers a spiral over a ring for these substructures in HT\,Lup\,A: solutions with $0^\circ$ pitch angle (i.e.\ a ring) are excluded at the $2.5\sigma$ level.
But even at this high angular resolution it is difficult to resolve the substructure in HT\,Lup\,A. A bar-like emission is observed connecting the two spiral arms with the inner disk, something that is not described by our models (Figure~\ref{fig:HTLup_spirals_bestfit}).

On the other hand, the AS\,205\,N disk has spiral features that are well resolved by our observations. These spirals appear to be better described by an Archimedean model than a logarithmic one, in terms of the reduced $\chi^2$ of each model, and since the former model better captures the spiral shape at both spiral ends. Thus, a variable pitch angle is preferred over a constant pitch angle to describe the \AS spirals. 

Interestingly, HT\,Lup\,A and AS\,205\,N show spiral features in the dust over a smaller range of radii and azimuthal angles than the other DSHARP spiral detections in \citet{Huang2018b}.
In fact, the radial extent of the spirals ($\approx36\,$au for \AS and $\approx4\,$au for HT\,Lup\,A) are smaller than the spirals observed in Elias\,27, IM\,Lup, and Wa\,Oph\,6, which range in radial extent from $\approx50\,$au to $\approx180\,$au. This difference in size could be explained by the truncation of the outer disk predicted in binary disks simulations or simulations of disks that have been involved in fly-by encounters 
\citep{Clarke1993,Breslau2014,Winter2018}
This will be further discussed in Section~\ref{sec:gas_discuss}.

\subsubsection{Axisymmetric substructures}

The bright rings and dark annulus observed in the single systems with spirals from DSHARP \citep[][]{Huang2018b} are not observed in the disks of the multiples systems presented here.
The lack of additional substructures may be explained by the small disk sizes: in the single systems with spiral arms, the gaps/rings usually show up at radial distances $\gtrsim 75\,$au, while our largest disk (AS\,205\,N) appears truncated at $\lesssim 60\,$au. 
If ring-like substructures are formed from pressure traps induced by planets, the lack of this substructures in \HT and \AS disks might suggest that stellar encounters and close binary companions inhibit planet formation, through disk truncation and material stripping, in agreement with the lower frequency of planet detections around binaries compared to single star systems \citep{wang2014}.

Only AS\,205\,S displays a prominent outer ring at $34\,$au, with a bright inner disk out to $\sim20$\,au, with both substructures separated by a gap. When imaged with a robust value of $-0.5$ we obtain an image of the disk with a smaller beam size ($16\,$mas, $2\,$au resolution at best), in which the inner disk is no longer centrally peaked, and a cavity starts to be resolved (see Figure~\ref{fig:AS205_mainfig}). This constrains the spectroscopic binary separation to be smaller than $\sim2$\,au in the southern component of AS\,205.

The unperturbed nature of the AS\,205\,S dust ring is puzzling. Assuming the spectroscopic binary has a total mass of $1.3\,M_\odot$ \citep{Eisner2005}, the orbital period at the ring location would be 173\,yr.
Close encounter simulations of star-disk interactions show that tidal stripping and arc-like features can be induced in timescales from hundred to few thousand years \citep[e.g. RW Aurigae, ][]{Dai2015}.
Thus, if the AS\,205 system had a fly-by interaction (see Section~\ref{sec:gas_discuss}) the ring at 34\,au in AS205S has only had $\lesssim 10$ orbits to recover its structure after the interaction. Most likely, the dynamical interaction could not have originated from a very close encounter, which would have disrupted or severely affected this ring.

\subsection{A fly-by in AS\,205 system? \label{sec:gas_discuss}} 

By analyzing the gas kinematics, disk rotation is identified in the N and S components of AS\,205. However, we also observe non-Keplerian features such as the flow or bridge of material between the disks, an extended arc-like emission (arc A) towards the north in AS\,205\,N, an asymmetric emission towards the South-East in AS\,205\,S, and a tilt/twist of the projected axis of rotation in AS\,205\,N and S (see moment 1 map of these object in Figure~\ref{fig:AS205_moms}). 

These features are quite similar to the ones observed in fly-by interactions in parabolic-like orbits. For example, the 3D hydrodynamic fly-by simulations from \cite{Dai2015} exhibit several of the non-Keplerian features present in the AS205 system, such as the gas bridge between the disks and the arc-like structure extending from the main component.
Since for prograde encounters (i.e.\ when the interaction occurs in the same direction as the disk rotation) the material stripped and/or transfered from one disk to the other is more pronounced, rather than for retrograde encounters \citep{Clarke1993,Dai2015}, it seems that the prograde case is a closer match to our observations.
In addition, an orthogonal or prograde non-coplanar encounter is expected to produce a warp in the disk \citep{Clarke1993, Cuello2018}, tilting and then twisting it, modifying the line of nodes, as observed in the velocity field of \AS and S (moment 1 maps, Figure~\ref{fig:AS205_moms}).

The truncated spatial extent of \AS in the dust can also be explained by a prograde fly-by interaction, which would strip off material from the main disk resulting in a outer disk radius that depends on the companion mass, the angle of interaction between disk plane and companion orbit, and the distance at periastron \citep{Clarke1993}. Further simulations and observations will help to better constrain this parameters.

\subsection{Disk misalignment in the HT\,Lup\,A-B binary}

In the close binary HT\,Lup\,A-B, we observe an apparent counter-rotation of their disks in Figure~\ref{fig:HTLup_mom1}. Given the degeneracy in the estimate of the misalignment between the disks, two different cases can explain the observations:
\begin{enumerate}
    \item The angular momentum vectors are misaligned by $91^\circ$, leaving the disks almost perpendicular to each other. If we assume that the spiral arms in \HT are trailing (i.e.\ the disk rotation is counter-clockwise) then the nearest side of the disk is in the East. For such a misalignment, assuming counter-clockwise rotation, the nearest side in HT\,Lup\,B would then be in the West, and the observed counter-rotation of the two disks would merely be a projection effect.
    Since we do not observe features in the CO channel maps that indicate transfer of material, an obvious perturbation in the butterfly pattern, or disk truncation in gas, this configuration is only possible if their physical separation is large enough in our line of sight.
    \item The angular momentum vectors are misaligned by $164^\circ$, with the disks close to parallel and counter rotating. As the trailing spirals assumption implies that \HT rotates counter-clockwise, HT\,Lup\,B would  in this case rotate clockwise, and its nearest side towards us would be its East side. As in the previous case, this would only be possible if the two disks are not in the same plane but instead, physically distant along our line of sight.
\end{enumerate}

Misalignments have been previously identified in other multiple systems \citep{Jensen2014, Williams2014, Brinch2016}, but the separations of these systems are much larger (hundreds of au) than in HT\,Lup\,A-B.  \cite{Bate2018} present hydrodynamical simulations that indicate that misaligned disks in binaries are possible, mainly due to fragmentation in turbulent environments and stellar capture. In addition, a complete flip of the disk orientation can occur due to gas accretion from the cloud with different angular momentum. These results show that in principle misalignments can arise in any direction depending on the cloud surroundings and environment where the disks are formed. A possible scenario for the formation of the HT\,Lup\,A-B system is an independent fragmentation of the binary components from the cloud, followed by a capture and subsequent orbital decay, leaving them close together and misaligned.

Future observations with high SNR in molecular lines, less contaminated from the cloud, and at similar or higher angular resolution, should be able to discern between the two scenarios, solving the degeneracy of disk orientation. A follow up of the HT\,Lup\,B and HT\,Lup\,C orbital positions will be also needed to get a complete description of the heavy truncation, misalignments, and dynamics in the HT\,Lup system.


\section{Conclusions}
\label{sec:conclusions}
We present very high resolution ($\sim 5\,$au scales) ALMA observations of the  multiple stellar systems AS\,205 and HT\,Lup, observed in band 6 (1.3mm) as part of DSHARP \citep{Andrews2018}.

In the continuum emission, the AS\,205 system resolves into two separate disks located at $168\,$au projected separation. The disk around AS\,205\,N shows two spiral arms extending from about 20 to $55\,$au in radius and over $180^{\circ}$ in azimuthal angle. By fitting them with an Archimedean and logarithmic spiral models, we find these arms to have similar pitch angles, close to $14^{\circ}$, although these are better described by the Archimedean model with a radially varying pitch angle.
The southern component, AS\,205\,S, displays an inner disk and a bright ring at $34\,$au, with a gap between the inner disk and outer ring that is not devoid of continuum emission.
The CO observations of AS\,205 show extended emission in the form of arc-like structures, with non-Keplerian motions observed around both disks. We found evidence to support that these features were triggered by the binary interaction via a close encounter or fly-by, which was also suggested by \cite{Salyk2014}. In this scenario, the AS\,205 system would have either a highly eccentric orbit between its components, or had a recent unbound interaction. Nevertheless, the regularity of the AS\,205\,S dust ring puts constrains over the proximity and timescale of this interaction, since the ring does not appear as perturbed as the gas.

For the first time at millimeter wavelengths, we resolve the two companions (B and C) in the HT\,Lup system, the closest one with a projected separation of 25\,au from HT\,Lup\,A. HT\,Lup\,C is located at more than 400\,au from HT\,Lup\,A-B. The disks around HT\,Lup\,B and C are the smallest objects in the DSHARP sample, with deconvolved sizes of $\approx 5$ and $\approx 10\,$au,  respectively. The HT\,Lup\,A disk is resolved and spiral structure is observed, which we constraint to be symmetric with a pitch angle close to $4^{\circ}$. However, the spirals are quite compact and appear to connect with the inner disk through a bar-like structure. Higher angular resolution might be needed in future observations to completely characterize this additional substructure. The kinematics of the CO emission in the closest binary, HT\,Lup\,A-B, shows an apparent counter-rotation of their disks. Given the degeneracy in disk orientation, we find two possible explanations depending on the angle between their angular momentum vectors, which could  either be a near to perpendicular relative orientation of their disks, in which case the counter-rotation would only be a projection effect, or alternatively, a close to parallel orientation of the disks with a physical counter-rotation that requires the disks to not be on the same plane.

The observations from DSHARP of multiple young stellar systems presented here, as well as future ALMA observations of gas and dust tracers at high angular resolution, are excellent laboratories to study dynamical interactions in multiple systems and to understand how this may affect the process of star and planet formation. This work is the first step towards a better understanding of how binary interactions and fly-bys affect disks structures, its evolution, and the efficiency of planet formation.

\bigskip

\acknowledgments \small
We are thankful to N.\ Cuello for insightful discussions. 
This paper makes use of ALMA datasets \newline \dataset[ADS/JAO.ALMA\#2016.1.00484.L]{https://almascience.nrao.edu/aq/?project\_code=2016.1.00484.L} and  \newline \dataset[ADS/JAO.ALMA\#2011.0.00531.S]{https://almascience.nrao.edu/aq/?project\_code=2011.0.00531.S}. ALMA is a partnership of ESO (representing its member states), NSF (USA) and NINS (Japan), together with NRC (Canada), MOST and ASIAA (Taiwan), and KASI (Republic of Korea), in cooperation with the Republic of Chile. The Joint ALMA Observatory is operated by ESO, AUI/NRAO and NAOJ.
Powered@NLHPC: This research was partially supported by the supercomputing infrastructure of the NLHPC (ECM-02), Center for Mathematical Modeling CMM, Universidad de Chile.
L.P. acknowledges support from CONICYT project Basal AFB-170002 and from FCFM/U. de Chile Fondo de Instalaci\'on Acad\'emica.
M.B. acknowledges funding from ANR of France under contract number ANR-16-CE31-0013 (Planet Forming disks).
Z. Z. and S. Z.acknowledges support from the National Aeronautics and Space Administration through the Astrophysics Theory Program with Grant No. NNX17AK40G and Sloan Research Fellowship. Simulations are carried out with the support from the Texas Advanced Computing Center (TACC) at The University of Texas at Austin through XSEDE grant TG- AST130002. 
J.H. acknowledges support from the National Science Foundation Graduate Research Fellowship under Grant No. DGE-1144152. 
S. A. and J. H. acknowledge funding support from the National Aeronautics and Space Administration under grant No. 17-XRP17\_2-0012 issued through the Exoplanets Research Program.
C.P.D. acknowledges support by the German Science Foundation (DFG) Research Unit FOR 2634, grants DU 414/22-1 and DU 414/23-1.
A.I. acknowledges support from the National Aeronautics and Space Administration under grant No. NNX15AB06G issued through the Origins of Solar Systems program, and from the National Science Foundation under grant No. AST-1715719.
V.V.G. and J.C acknowledge support from the National Aeronautics and Space Administration under grant No.\ 15XRP15\_20140 issued through the Exoplanets Research Program.
L. R. acknowledges support from the ngVLA Community Studies program, coordinated by the National Radio Astronomy Observatory, which is a facility of the National Science Foundation operated under cooperative agreement by Associated Universities, Inc.



\appendix

\section{Figures \label{app:figures}}

\begin{figure*}[ht!]
\includegraphics[width=\linewidth]{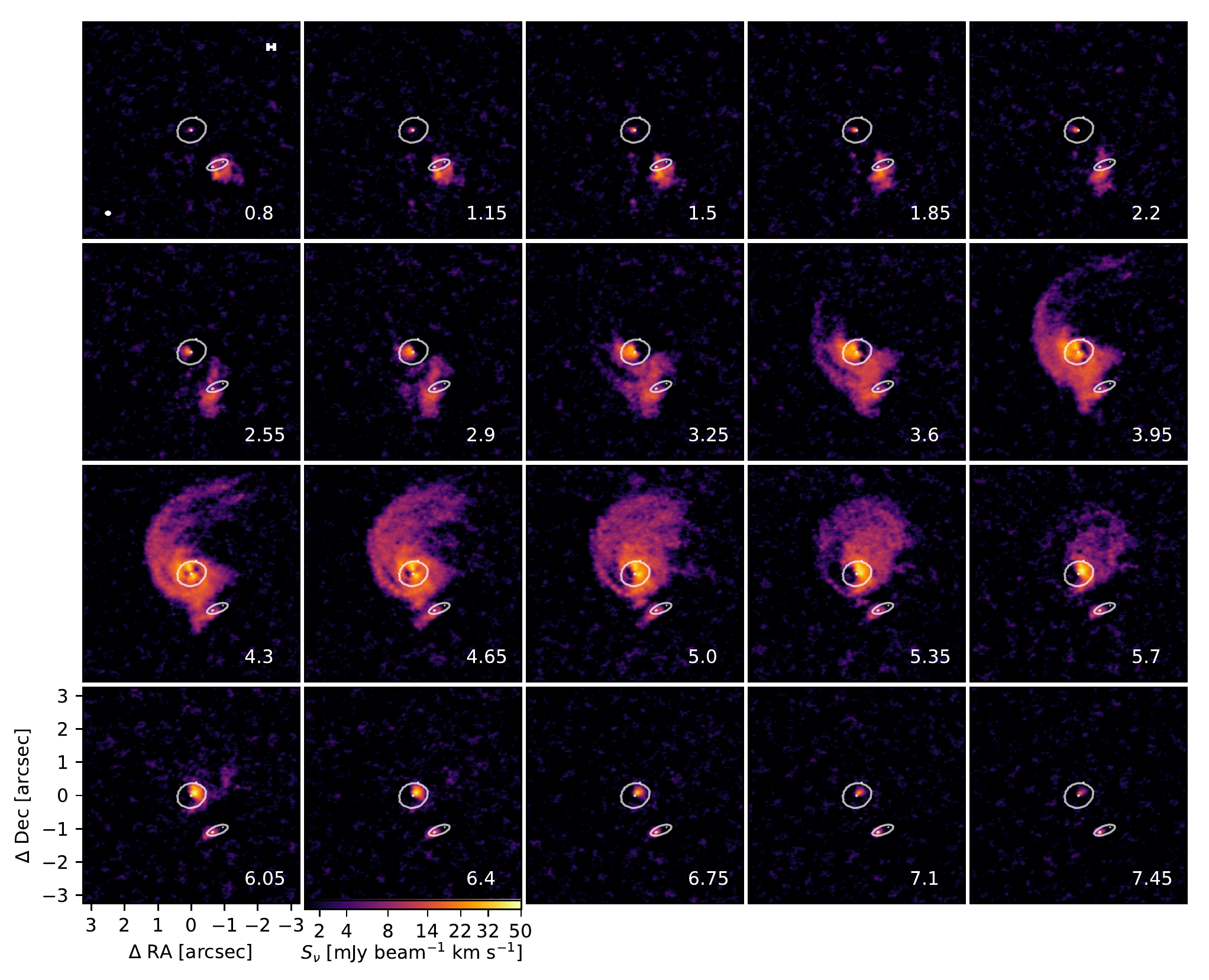}
\caption{Channel maps of the CO emission in the AS\,205 system. Each box is $7\farcs{5}$ wide, centered at the continuum luminosity peak of AS\,205\,N. The LSRK velocity is printed at the lower right corner of each frame, while the beam size and a $25\,$au scale bar are printed in the first image at the upper left frame. The contour levels correspond to $5\sigma$ and $300\sigma$ in continuum, for comparison.
\label{fig:AS205_CO_gallery}}
\end{figure*}

\begin{figure*}[ht!]
\includegraphics[width=\linewidth]{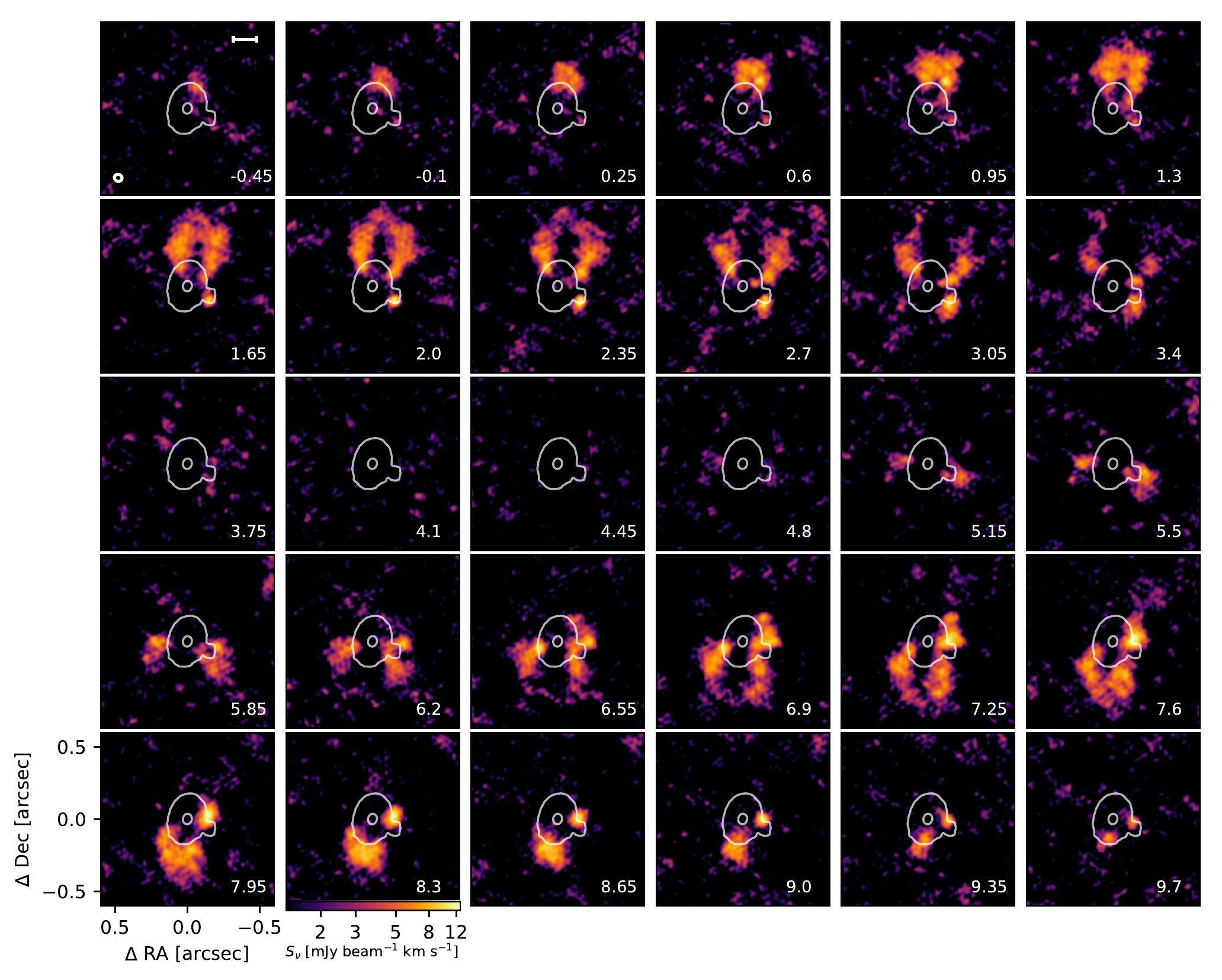}
\caption{The CO emission of HT\,Lup\,A/HT\,Lup\,B. Each square is $1\farcs{2}$ per side, and centered at the peak flux of HT\,Lup\,A in continuum. The LSRK velocity is printed at the lower right corner of each frame, while the beam size and a $25\,$au scale bar are printed in the first image at the upper left frame. The image was generated by removing all baselines with less than $150\,$m in length, losing sensitivity at scales larger than $2\arcsec$. The contour levels correspond to $5\sigma$ and $300\sigma$ in the continuum image, for comparison.
\label{fig:HTLup_CO_gallery}}
\end{figure*}

Here we present the spectral data cubes (channel maps) of AS\,205 (Figure \ref{fig:AS205_CO_gallery}) and HT\,Lup (Figure \ref{fig:HTLup_CO_gallery}).

\end{document}